\documentclass[aps,prc,preprint,superscriptaddress]{revtex4}

\usepackage{graphicx}
\usepackage{booktabs}
\usepackage{hyperref}
\usepackage{amssymb,amsmath,amsfonts}
\usepackage{slashed}

\begin{document}


\title{Dense Baryonic Matter and Strangeness in Neutron Stars\footnote{Invited talk at the {\it 8th International Conference on Quarks and Nuclear Physics} (QNP 2018), Tsukuba, Japan. To appear in {\it JPS Conf. Proc.}}}

\author{Wolfram Weise}
\affiliation{Physics Department, Technical University of Munich, 85748 Garching, Germany}

\vspace{5cm}

\begin{abstract}
Recent developments of chiral effective field theory (ChEFT) applications to nuclear and neutron matter are summarized, with special emphasis on a (non-perturbative) extension using functional renormalisation group methods. Topics include: nuclear thermodynamics, extrapolations to dense baryonic matter and constraints from neutron star observables. Hyperon-nuclear interactions derived from SU(3) will be discussed with reference to the ``hyperon puzzle" in neutron star matter.
\end{abstract}



\maketitle

\section{Introduction}

Understanding nuclei and dense baryonic matter at the interface with QCD is one of the pending challenges in nuclear many-body theory. The present paper reports on recent developments using effective field theory methods applied to dense baryonic matter.  Of special interest in this context is the equation-of-state (EoS) of cold and dense matter subject to constraints from neutron star observations (the existence of two-solar-mass stars \cite{Demorest2010, Fonseca2016, Antoniadis2013} and information deduced from gravitational wave signals of two merging neutron stars\cite{Abbott2017,Most2018} ). Whatever the detailed composition of matter in the core of neutron stars may be (see also the presentation by G. Baym at QNP2018), its EoS must produce sufficiently high pressure in order to meet the empirical benchmark constraints from astrophysics together with those from nuclear physics at lower densities. A further persisting question in this context is about the possible presence of strange quarks or hyperons in neutron stars. This issue, the so-called ``hyperon puzzle", will be addressed in a later section of this presentation. 

\section{Nucleons and pions in low-energy QCD}

Global symmetries of QCD provide guidance for constructing effective field theories that represent low-energy QCD and thus establish a conceptual frame for nuclear physics and extrapolations to matter at higher densities. In its sector with the two lightest ($u$ and $d$) quarks, {\it chiral symmetry} is a key to understanding low-energy energy hadron structure as well as nuclear forces. Given the small quark masses ($m_u\simeq 2$ MeV and $m_d\simeq 5$ MeV at a renormalization scale $\mu\simeq 2$ GeV), a useful starting point is QCD with a massless isospin doublet $(u,d)$ of quarks. In this limit, chiral invariance is an {\it exact} symmetry of QCD. Non-perturbative quark-gluon dynamics implies that this symmetry is {\it spontaneously} broken at low energy. The Nambu-Goldstone (NG) boson of spontaneously broken chiral symmetry is identified as the isospin triplet of pions.

Starting from three massless quarks forming a color-singlet, the massive nucleon is generated by quark-gluon dynamics. The three valence quarks, accompanied by a strong gluonic field and a sea of quark-antiquark pairs, are localized in a small volume with a radius of less than a Fermi. Localization (confinement) of the valence quarks in a finite volume implies breaking of chiral symmetry. The following low-energy QCD based picture of the nucleon, relevant to nuclear physics, thus emerges. The underlying strong-interaction processes which generate the nucleon mass $M_N$ induce at the same time the spontaneous breaking of chiral symmetry. The three valence quarks, carrying one unit of baryon number, form a compact core of the nucleon. Strong vacuum polarization effects surround this baryonic core with multiple quark-antiquark pairs. The low-energy physics associated with this mesonic ($\bar{q}q$) cloud is governed by pions as NG bosons of spontaneously broken chiral symmetry, in such a way that the total axial vector current of the ``core + cloud" system is conserved apart from small mass corrections (PCAC). 

Interactions between two nucleons at long and intermediate distances are generated as an inward-bound hierarchy of one-, two-, multi-pion exchange processes. The framework for systematically organizing this hierarchy is chiral effective field theory (ChEFT). It is based on the separation of scales delineating low energies and momenta characteristic of nuclear physics from the chiral symmetry breaking ``gap" scale, $\Lambda_\chi = 4\pi f_\pi \sim 1$ GeV $\sim M_N$. In this context the pion decay constant, $f_\pi \simeq 0.09$ GeV, figures as an order parameter for spontaneously broken chiral symmetry.

\section{From nucleons to compressed baryonic matter}

Extrapolations from normal nuclear systems to the high-density regime require a detailed assessment of size scales of the nucleon itself.  As mentioned the ``chiral" nucleon is viewed as a compact valence quark core surrounded by a multi-pion cloud. Early descriptions of the nucleon as a topological soliton derived from a non-linear chiral meson Lagrangian \cite{KMW1987} suggested an r.m.s. radius of about 0.5 fm for the baryonic core, together with a significantly larger charge radius determined primarily by the charged meson cloud surrounding the core. The ratio of the corresponding volumes of baryon no. and isoscalar charge distributions, $V_{baryon \,no.}$ / $V_{charge} \sim 0.2$, indicates yet another separation of scales relevant to the window of applicability for ChEFT.

Empirical tests delineating the size scale of the nucleon core begin to be accessible in deeply-virtual Compton scattering experiments at J-Lab and their theoretical analysis \cite{CLAS2015a, CLAS2015b, DGV2017}. An interesting development in this respect is the measurement of form factors of the nucleon's energy momentum tensor (see also the presentation by B. Pasquini at QNP2018). First explorations of the distribution of pressure inside the proton \cite{BEG2018} appear to be consistent with previous chiral quark-soliton model
predictions \cite{PS2018, Goeke2007} and the picture of a compact baryonic core surrounded by a pionic cloud. 

\begin{figure}[th]
\centerline{\includegraphics[width=10cm]{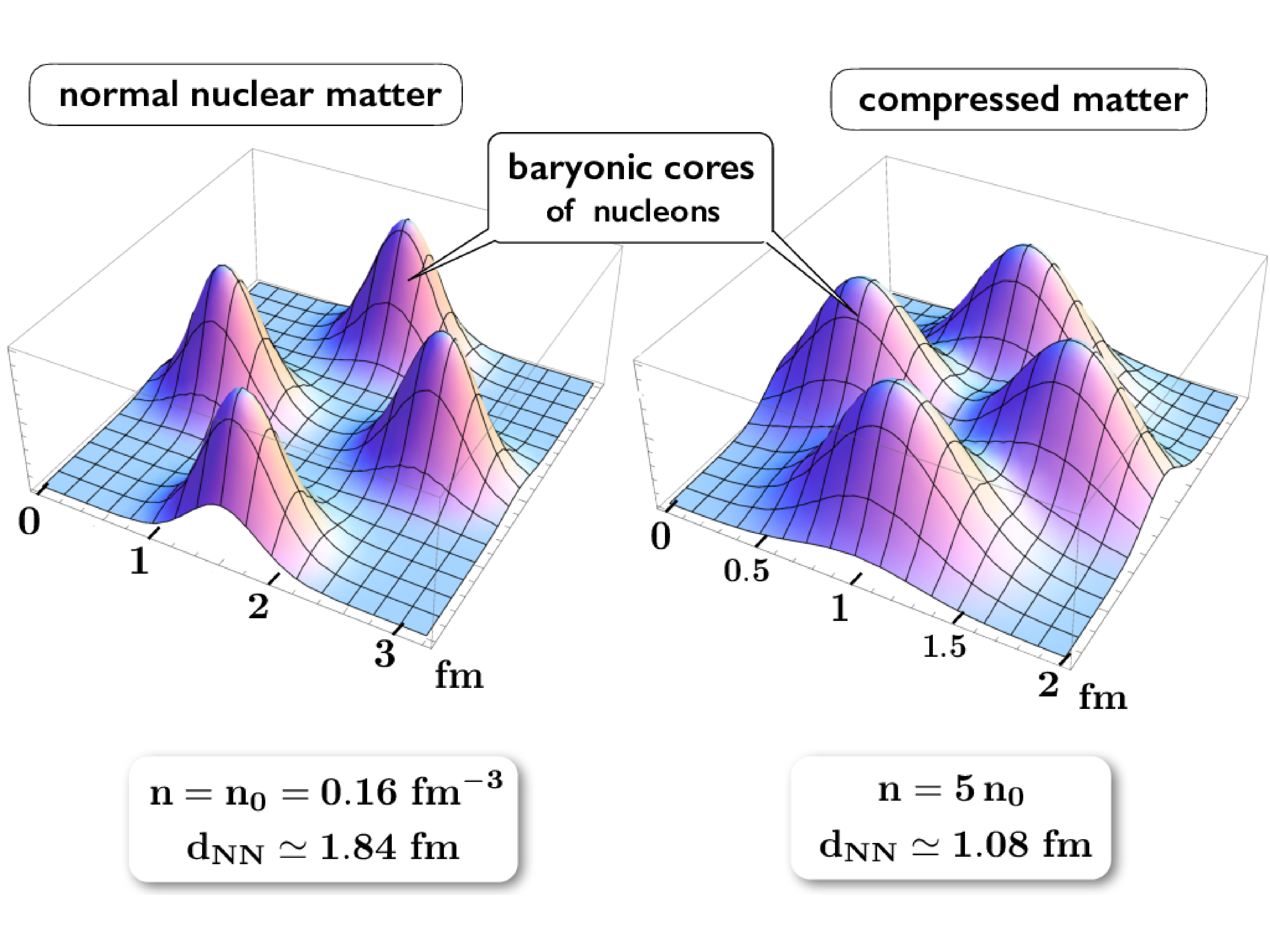}}
\caption{Schematic picture of baryonic matter at normal nuclear density ($n_0 = 0.16$ fm$^{-3}$, left) and at $n = 5\,n_0$ (right). The individual baryon number density distributions are Gaussians with an r.m.s. radius $\sqrt{\langle r_B^2\rangle}=0.5$ fm.}
\label{fig:1}
\end{figure}

Assuming a typical 1/2-fm radius of the baryon core, let us consider compressed baryonic matter and examine up to which baryon densities, $n = B/V$, one can still expect nucleons (rather than free-floating quarks) to be the relevant baryonic degrees of freedom. A schematic picture is drawn in Fig.\,\ref{fig:1}, illustrating a piece of baryonic matter as a set of Gaussian distributions. Each of these Gaussians approximates very well the baryon density of the nucleon as a chiral soliton. At the density of normal nuclear matter, $n_0 = 0.16$ fm$^{-3}$, the baryonic cores are well separated by an average distance $d_{NN} \sim n_0^{-1/3}\simeq 1.8$ fm. Pions couple to these baryonic sources and act in the space between these cores. The pion field incorporates multiple exchanges of pions between nucleons, and those mechanisms are properly dealt with in chiral EFT.

While the longer-range pionic field configurations will undergo drastic changes in highly compressed baryonic matter, the sizes of the compact baryonic cores themselves are expected to remain stable as long as they continue to be separated. As illustrated in Fig.\,\ref{fig:1}, even at $n = 5\,n_0$, corresponding to an average distance $d_{NN} \simeq 1.1$ fm between nucleons, the individual baryon distributions are indeed still well identifiable, with just small overlaps at their touching surfaces. At such high densities the (non-linear) pion field between baryonic sources is accumulating much strength. Non-perturbative chiral field theory methods must be employed to treat these strong-field configurations.

In this picture the nucleons start loosing their identities once the density reaches $n\gtrsim 8\,n_0$ and the baryon distributions begin to merge (percolate). Quark matter with strong pairing (diquarks) is supposed to take over\cite{Baym2018}. However, such extreme densities are already beyond the baryon densities typically encountered in the central regions of neutron stars if their radii are larger than about ten kilometers. 
Furthermore the strong short-distance repulsion between nucleons makes a ``soft" merging scenario for nucleons energetically quite expensive. The repulsive hard core of the $NN$ interaction with its typical range of about half a Fermi has a long phenomenological history in nuclear physics. More recently this hard core has been established by deducing an equivalent local $NN$ potential from lattice QCD computations\cite{AHI2010,Aoki2013} .

\section{Nuclear chiral effective field theory}

Chiral effective field theory starts out as a non-linear theory of pions and their interactions, with symmetry-breaking mass terms added. Nucleons are introduced as ``heavy" sources of the NG bosons. A systematically organized low-energy expansion in powers of derivatives of the pion field (power-counting) provides a remarkably successful quantitative description (chiral perturbation theory) of pion-pion scattering and of pion-nucleon interactions at momenta and energies small compared to the chiral symmetry breaking scale, $\Lambda_\chi\sim 4\pi f_\pi$. ChEFT with inclusion of nucleons is the basis for a successful theory of the nucleon-nucleon interaction. In recent years this theory has also become the widely accepted framework for the treatment of nuclear many-body systems (see e.g. \cite{HKW2013, HRW2016} and references therein for recent overviews). 

Nuclear forces in ChEFT are constructed in terms of a systematically organized hierarchy of explicit one-, two- and
multi-pion exchange processes, constrained by chiral symmetry, plus a complete set of contact terms
encoding unresolved short-distance dynamics \cite{evgenireview,hammerreview,machleidtreview}. Three-body forces enter at next-to-next-to-leading order ($N^2LO$) in this hierachy and turn out to be important in reproducing the equilibrium properties of nuclear matter.

Chiral EFT is basically a perturbative framework. In a nuclear medium, the new ``small" scale that enters in addition to three-momenta and pion mass is the Fermi momentum, $p_F$, of the nucleons. Its value at the equilibrium density of $N = Z$ nuclear matter, $n_0 = 2 p_F^3/3\pi^2 = 0.16$ fm$^{-3}$, is $p_F = 1.33$ fm$^{-1} \simeq 1.9\, m_\pi$, small compared to the chiral scale $\Lambda_\chi = 4\pi f_\pi \sim 1$ GeV. Expansions in powers of $p_F/\Lambda_\chi < 0.3$ are thus likely to converge. Even at densities $n = 3 n_0$ the Fermi momentum still satisfies $p_F/\Lambda_\chi < 0.4$. These observations suggest indeed the applicability, within limits, of perturbative approaches to the nuclear many-body problem. 

\begin{figure}
\centerline{\includegraphics[width=8cm] {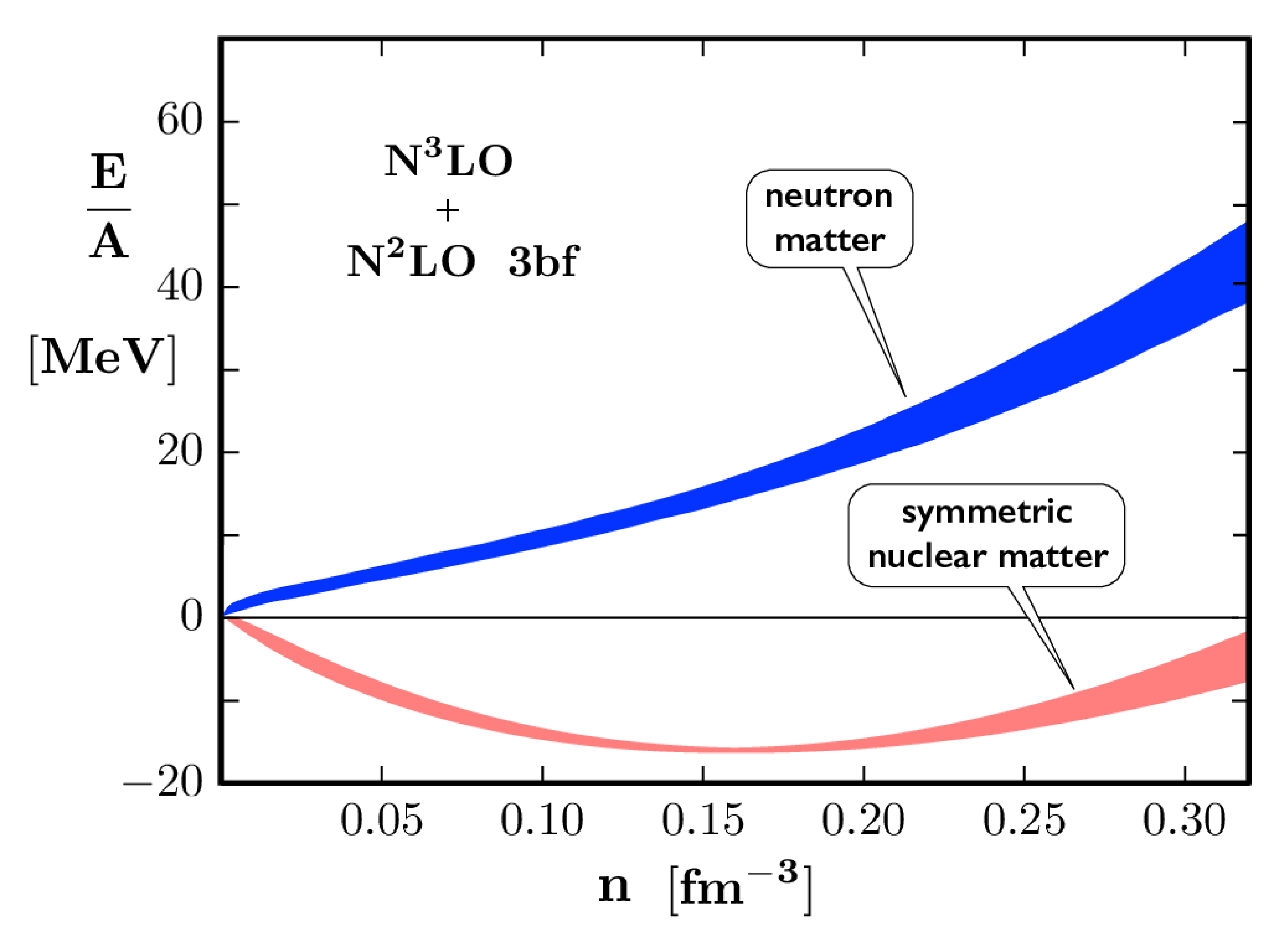}}
\caption{Energy per particle of symmetric nuclear matter and neutron matter at temperature $T=0$ as function of baryon density. Results of ChEFT calculations using $N^3LO$ chiral NN interactions and $N^2LO$ three-body forces.  Adapted from ref.\cite{LH2018} .}
\label{fig:2}
\end{figure}

Examples of recent perturbative ChEFT computations of the energy per particle, $E/A$, for nuclear and neutron matter at zero temperature $T = 0$ are shown in Fig.\,\ref{fig:2}. Starting from $N^3LO$ chiral $NN$ interactions and $N^2LO$ three-body forces, calculations using third-order many-body perturbation theory have been performed \cite{HK2017,LH2018} . These calculations are considered reliable up to densities $n \sim 2\,n_0$ and perhaps slightly beyond, but because of their perturbative nature, they cannot be extended up to the densities relevant for neutron star cores. 

\section{Non-perturbative strategies: functional renormalization group}

The perturbative chiral EFT approach to nuclear and neutron matter relies on a convergent expansion of the energy-density in powers of $p_F/\Lambda$, the ratio of Fermi momentum and a suitably chosen momentum space cutoff. Typical ChEFT cutoffs are $\Lambda \sim 400-500$ MeV, about half of the chiral symmetry breaking scale $\Lambda_\chi = 4\pi f_\pi$.  At densities $n \sim 5 n_0$ the Fermi momenta are comparable to $\Lambda$ and therefore a perturbative expansion cannot be expected to work. A non-perturbative method needs to be developed. In particular, multi-nucleon correlations grow rapidly with increasing density and must be treated through proper resummations. 

Here our method of choice is the functional renormalization group (FRG) \cite{We1993} applied to nuclear many-body systems. We give a brief outline and summary of developments and results reported in \cite{DW2017,DW2015,DHKW2013}. The underlying logic is the following: in the domain of spontaneously broken chiral symmetry, the active degrees of freedom are pions and nucleons. The ``ultraviolet" (UV) initialization of functional renormalization group (FRG) flow equations \cite{We1993} starts at the chiral symmetry breaking scale, $\Lambda_\chi = 4\pi f_\pi$. The FRG evolution of the action proceeds down to low-energy ``infrared" (IR) scales characteristic of the nuclear many-body problem at Fermi momenta $p_F\ll  \Lambda_\chi$. At the UV scale a chiral nucleon-meson Lagrangian based on a linear sigma model with an appropriate (non-linear) effective potential and short-distance vector-current couplings is chosen as a starting point. This chiral nucleon-meson field theory involves a scalar $\sigma$ field accompanying the pion as a chiral partner. The primary $\sigma$ is heavy, with a mass close to 1 GeV reminiscent of the $f_0(980)$, and not to be confused with the broad ``$\sigma(500)$" which is generated dynamically as a pole in the s-wave $\pi\pi$ scattering T -matrix. 

The effective action in the IR limit then emerges by solving the non-perturbative FRG flow equations. As a consistency condition, the physics results in this long-wavelength, low-density limit should match those from perturbative chiral effective field theory. The strength of the chiral FRG approach is that it incorporates resummations to all orders of important multi-pion fluctuations, nucleon-hole excitations (i.e. fluctuations around the nuclear Fermi surface) and many-body correlations. It can therefore be extended up to high baryon densities as long as the system remains in the spontaneously broken (Nambu-Goldstone) realisation of chiral symmetry.

This non-perturbative FRG framework based on chiral nucleon-meson field theory yields results for symmetric and asymmetric nuclear matter as well as pure neutron matter that are consistent with those of perturbative chiral EFT at moderate temperatures and densities. In particular, nuclear thermodynamics including the liquid-gas phase transition is well reproduced as discussed in more detail in ref.\,\cite{DW2017}.

\section{Chiral symmetry restoration and order parameters}

Whereas the applicability of perturbative chiral EFT is limited to baryon densities $n \lesssim 2\,n_0$, the non-perturbative chiral FRG approach can in principle be extended to compressed baryonic matter at higher densities.
A necessary condition for this to work is that matter remains in the hadronic phase characterized by the spontaneously broken Nambu-Goldstone realisation of chiral symmetry. 

Lattice QCD computations\cite{Borsanyi2010, Bazavov2012} at $\mu = 0$ demonstrate the existence of a crossover transition towards restoration of chiral symmetry in its Wigner-Weyl realisation at temperatures $T > T_c \simeq 0.15$ GeV. Chiral symmetry is presumably also restored at large baryon chemical potentials $\mu$ and low temperature. However, the critical value of $\mu$ at which this transition might take place is unknown. 

Several calculations of isospin-symmetric matter using Nambu \& Jona-Lasinio or chiral quark-meson models have predicted a first-order chiral phase transition at vanishing temperature for quark chemical potentials, $\mu_q$, around 300\,MeV (see, e.g., Refs.\cite{Fukushima2008,RRW2007,Schaefer2007,HRCW2009,Herbst2013}). Translated into baryonic chemical potentials, $\mu\simeq3\mu_q$, chiral symmetry would then be restored not far from the equilibrium point of normal nuclear matter, $\mu_0 =923$ MeV. Nuclear physics with its well-established empirical phenomenology teaches us that this can obviously not be the case. However, these calculations -- apart from the fact that they operate with (quark) degrees of freedom that are not appropriate for dealing with the hadronic phase of QCD -- work mostly within the mean-field approximation. It is therefore of great importance to examine how fluctuations beyond mean-field can change this scenario. Using chiral nucleon-meson field theory in combinaton with FRG, we observe indeed that the mean-field approximation cannot be trusted: it is likely that fluctuations, such as repulsive multi-nucleon correlations in the presence of the Pauli principle, shift the chiral transition to extremely high baryon densities beyond six times $n_0$.

\begin{figure}
	 \centerline{\includegraphics[width=8cm] {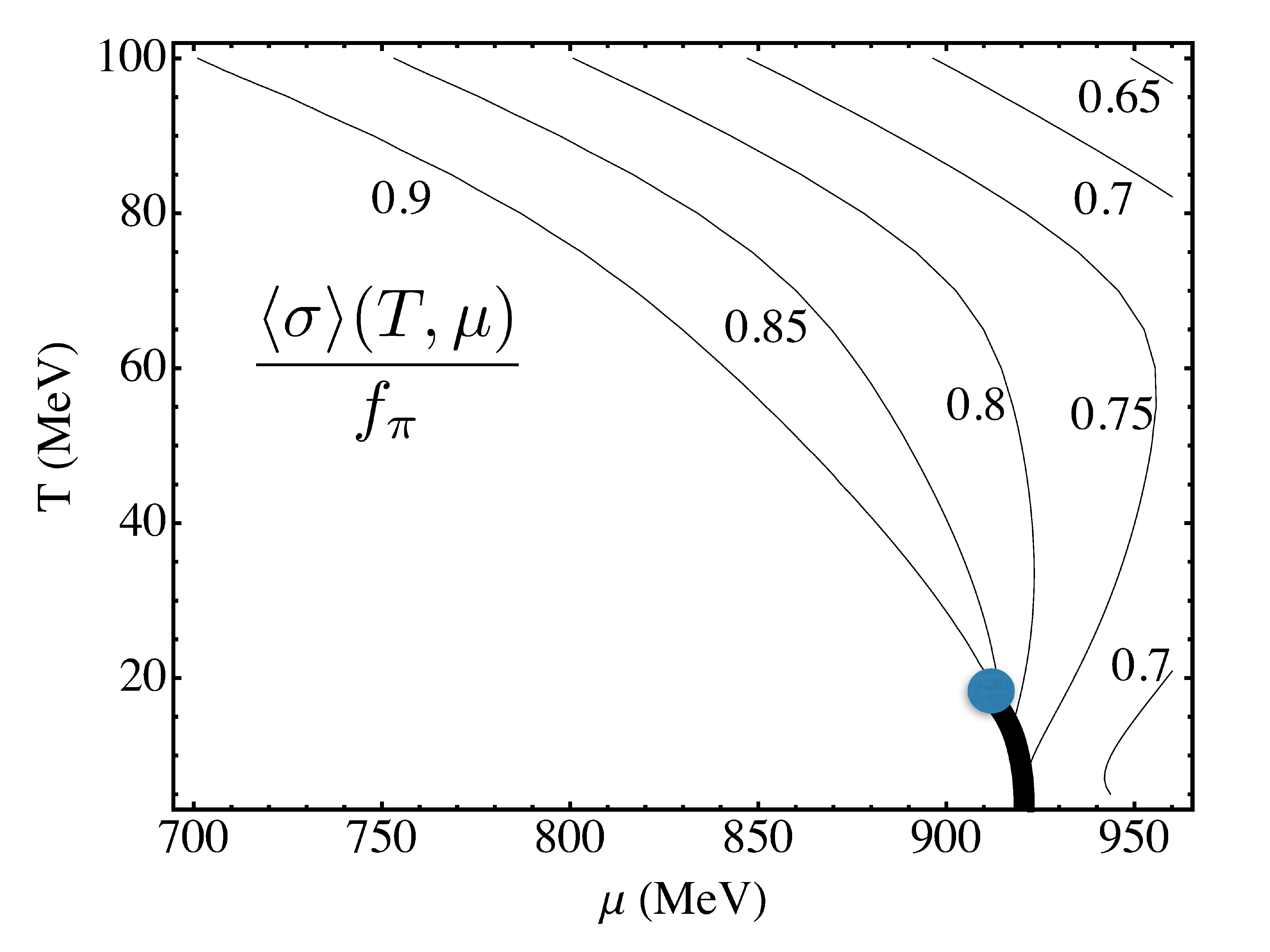}}
	\caption{$T-\mu$ phase diagram of symmetric nuclear matter calculated in the FRG-ChNM model \cite{DHKW2013}. Contour lines for constant values of the chiral order parameter $\langle\sigma\rangle/f_\pi$ are drawn with numbers attached. The liquid-gas first-order transition line and its critical point are also indicated for orientation.}
	\label{fig:3}
\end{figure}

In chiral nucleon-meson (ChNM) field theory with its chiral field $(\boldsymbol{\pi},\sigma)$, the expectation value of the scalar field, $\langle\sigma\rangle(\mu,T) = f_\pi^*(\mu,T)$, or equvalently, the in-medium pion decay constant, acts as order parameter for the spontaneous breaking of chiral symmetry. 
It is instructive to examine this order parameter in the $T$--$\mu$ phase diagram of symmetric nuclear matter around the liquid-gas phase transition. Figure\,\ref{fig:3} shows contours of constant values $\langle\sigma\rangle/f_\pi$ in the $T-\mu$ plane, calculated using the chiral nucleon-meson theory in combination with full FRG including fluctuations. Evidently, within the whole region of temperatures $T\lesssim 100$ MeV and baryon chemical potentials $\mu\lesssim 1$ GeV, the order parameter remains far from zero and there is no tendency towards a chiral phase transition. Nowhere in this whole $(T,\mu)$ range does the effective nucleon mass in the medium drop below $M^*_N(T,\mu) \simeq 0.7\,M_N$. 

\begin{figure}
	 \centerline{\includegraphics[width=8cm] {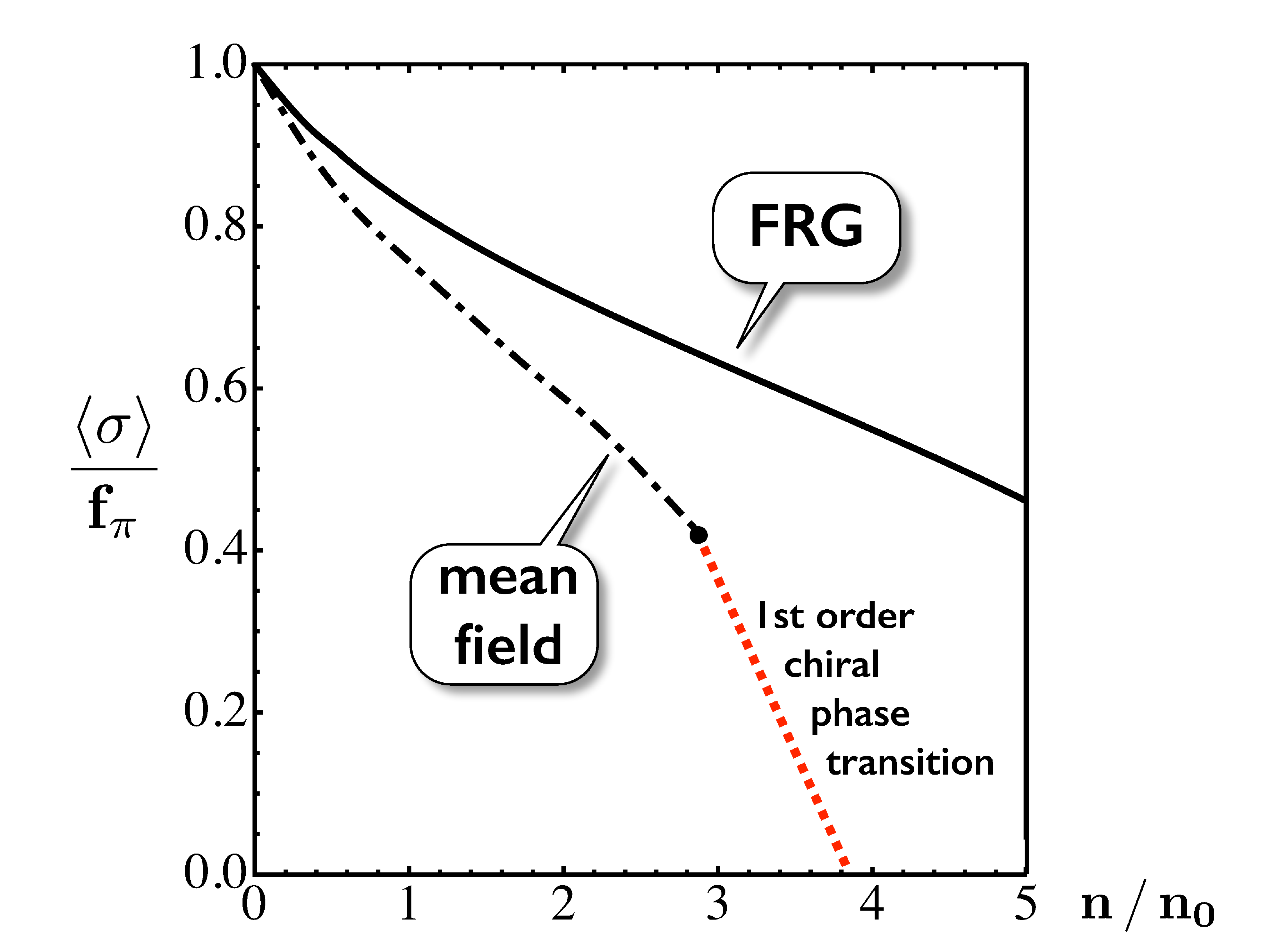}}
	\caption{Density dependence of the chiral order parameter for pure neutron matter at vanishing temperature\cite{DW2015}. Solid line (FRG): FRG calculation with chiral nucleon-meson theory input, including fluctuations. Dashed line (MF): mean-field approximation result featuring a first-order chiral phase transition.}
	\label{fig:4}
\end{figure}

For pure neutron matter at $T=0$, the chiral condensate has been calculated previously within (perturbative) chiral effective field theory\cite{Kaiser2009,Krueger2013} . Chiral nuclear forces treated up to four-body interactions\cite{Krueger2013} at N$^3$LO were shown to work moderately against the leading linearly decreasing chiral condensate with increasing density around and beyond $n \simeq n_0$. The non-perturbative FRG approach permits an extrapolation to higher densities.  Results are presented in Fig.\,\ref{fig:4}. In mean-field approximation the order parameter $\langle\sigma\rangle/f_\pi$ shows a first-order chiral phase transition at a density of about $3\,n_0$. However, the situation changes qualitatively when fluctuations are included using the FRG framework. The chiral order parameter now turns into a continuous function of density, with no indication of a phase transition. Even at five to six times nuclear saturation density the order parameter still remains at about forty percent of its vacuum value. Only at densities as large as $n \sim 7n_0$ does the expectation value of $\sigma$ show a more rapid tendency of a crossover towards restoration of chiral symmetry. But this is already beyond the range of densities that may be reached in the cores of neutron stars. 

As a significant outcome of the non-perturbative FRG computations, we thus observe a strong influence of higher order fluctuations involving Pauli blocking effects in multiple pion-exchange processes and multi-nucleon correlations at high densities. With neutron matter remaining in a phase of spontaneously broken chiral symmetry even up to densities as high as $6\,n_0$, this encourages further-reaching applications and tests of the FRG-ChNM approach in constructing an equation-of-state for the interior of neutron stars. 

\section{Equation-of-state for neutron star matter}

\begin{figure}
	 \centerline{\includegraphics[width=10cm] {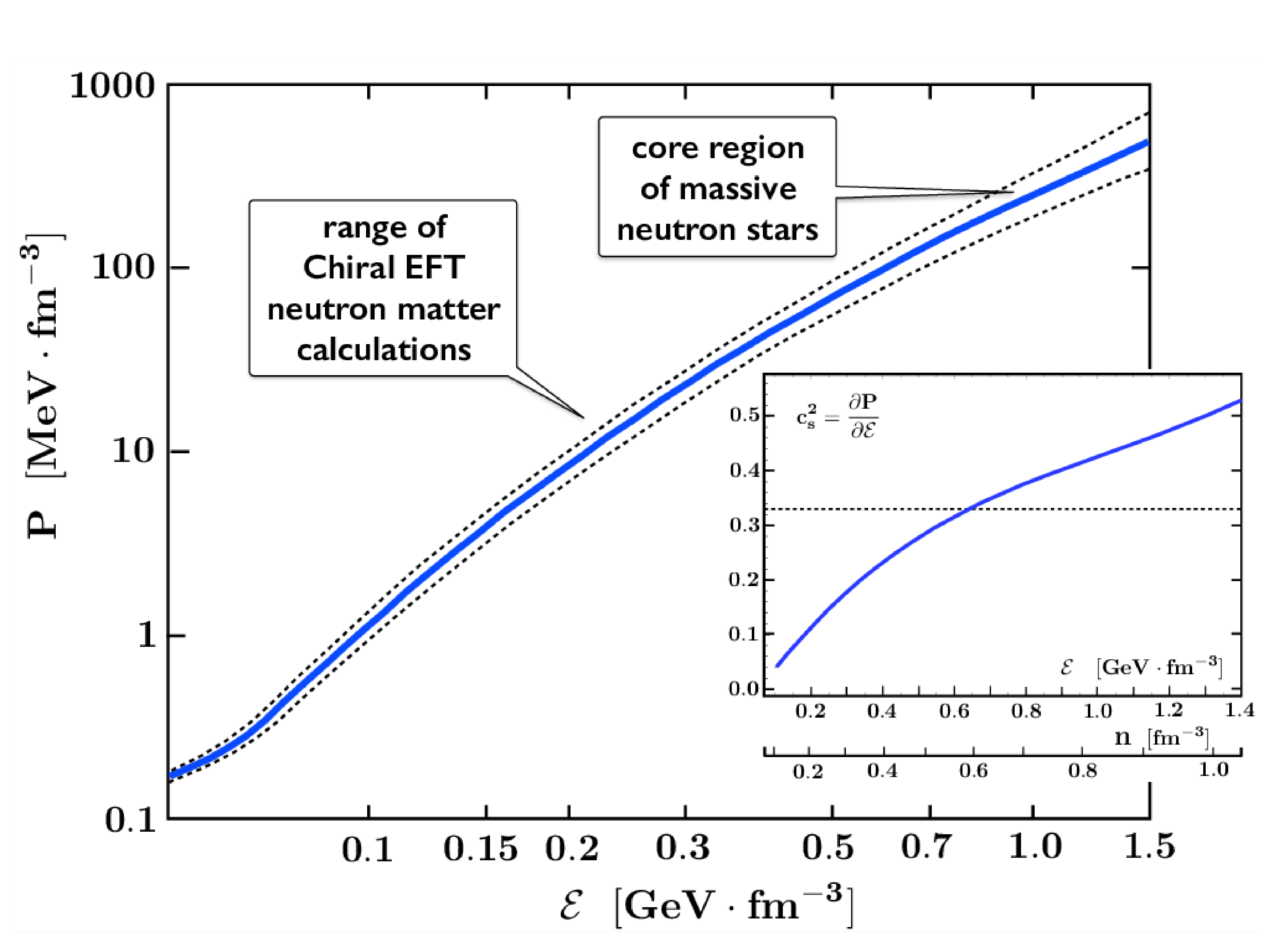}}
	\caption{Pressure as function of energy density fo neutron star matter with inclusion of beta equilibrium. Solid curve: result of a calculation\cite{DW2017,DW2015} starting from chiral nucleon-meson theory and solving functional renormalization group equations. The uncertainty band indicated by dotted lines corresponds to a range of input symmetry energy values 30 - 34 MeV at density $n = n_0 = 0.16$ fm$^{-3}$. Also shown is the square of the sound velocity, $c_s^2 = \partial P({\cal E})/\partial{\cal E}$.}
	\label{fig:5}
\end{figure}

Any equation-of-state for neutron star matter at $T=0$ must satisfy the aforementioned constraints (i.e. it must support  $2\,M_{\odot}$ stars and satisfy the tidal deformabillity limits deduced from neutron star merger events). An EoS which fulfills these conditions has been computed \cite{DW2015,DW2017} solving the FRG equations with input from chiral nucleon-meson theory. At low densities this EoS is consistent with chiral EFT calculations of neutron matter and symmetric nuclear matter. The calculated symmetry energy at $n = n_0$ is $[E(Z=0) - E(Z=N)]/A = 32$ MeV. Beta equilibrium for neutron star matter is properly incorporated. The result for the pressure as function of the energy density, $P({\cal E})$, in Fig.\,\ref{fig:5} shows a steep rise towards pressures exceeding 100 MeV/fm$^{3}$ in the region relevant for the core of heavy neutron stars. Such an amount of pressure is indeed capable of supporting a two-solar-mass neutron star. Its radius is predicted to be $R\simeq 11.5$ km. Notably the baryon density in the center of such an object does not exceed $n\sim 5\,n_0$. Following the previous discussions it then appears justified to work with nucleons and pion fields as relevant degrees of freedom even at such extreme but not hyperdense conditions.

A further interesting property of compressed baryonic matter is its velocity of sound. For a non-interacting relativistic Fermi gas the squared sound velocity has a canonical value, $c_s^2= {\partial P({\cal E})\over \partial{\cal E}} = {1\over 3}$, which is supposed to be reached at asymptotically high densities. The inset of Fig.\,\ref{fig:5} shows that the squared sound velocity of the FRG - ChNM equation-of-state exceeds $c_s^2 = 1/3$ at a baryon density around $n\sim 4\,n_0$ and continues to grow as it approaches neutron star core densities. This behaviour can be traced to the continuously rising strength of repulsive many-body correlations driven in part by the Pauli principle as the density increases. At much higher densities, once nucleon clusters dissolve and quark matter takes over, $c_s^2$ is expected to decrease again and ultimately approach its asymptotic value of 1/3 from below at ultrahigh densities \cite{TCGR2018} . 

\begin{figure}
	 \centerline{\includegraphics[width=10cm] {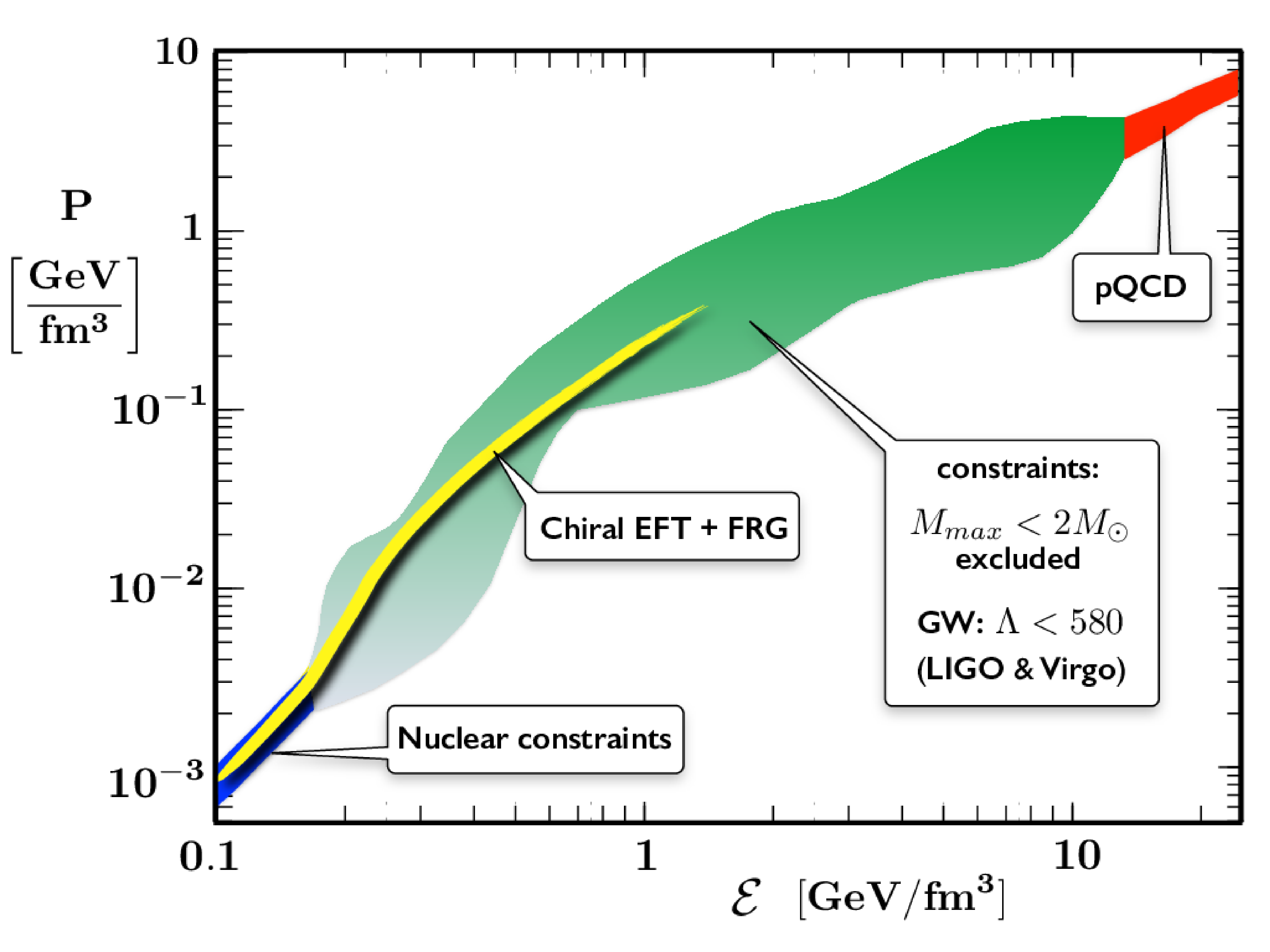}}
	\caption{Pressure as function of energy density for neutron star matter with constraints from perturbative QCD and nuclear EFT calculations at the upper and lower ends of the density scale, adapted from refs.\cite{Annala2018,Vuorinen2018}. The shaded area represents the range of acceptable equations-of-state subject to empirical conditions on neutron star maximum mass and tidal deformability, the latter from LIGO \& Virgo gravitational wave analysis. The curve denoted ``Chiral EFT + FRG" represents $P({\cal E})$ from Fig.\,\ref{fig:5}. }
	\label{fig:6}
\end{figure}

An important effort presently pursued and steadily being improved is to constrain the neutron star equation-of-state systematically from observational data, together with an interpolation between theoretical limits provided by nuclear physics at low densities and perturbative QCD at asymptotically high densities. Fig.\,\ref{fig:6} shows a recent example. Nuclear constraints as represented by ChEFT calculations\cite{HKW2013,HLPS2013} set the low-density limit of $P({\cal E})$ at energy densities ${\cal E}\lesssim 200$ MeV/fm$^3$. Sophisticated perturbative QCD calculations \cite{Kurkela2014} determine the pressure at extreme energy densities, ${\cal E} > 10$ GeV/fm$^3$. Constraints in the region between these extremes are introduced by studying large sets of parametrized equations-of-state subject to the condition that they all produce a maximum neutron star mass of at least $2\,M_\odot$ and at the same time fulfill the tidal deformability limit deduced from the updated LIGO \& Virgo gravitational wave analysis. The shaded area in Fig.\,\ref{fig:6} defines the region of acceptable neutron star equations-of-state under such conditions\cite{Annala2018, Vuorinen2018}. Remarkably, the EoS computed using the FRG-ChNM approach and shown in Fig.\,\ref{fig:5} satisfies these constraints up to the densities relevant for the core of massive neutron stars. Of course, one order of magnitude in the pressure $P$ still separates this neutron star domain from the perturbative QCD limit. It is in the range of densities $n > 5\,n_0$ where one can expect a (probably continuous) hadrons-to-quarks transition\cite{Baym2018} to take place.

\section{Strangeness in neutron stars ?}

An issue that still needs to be resolved is the so-called ``hyperon puzzle" in neutron stars. At densities around 2-3 times $n_0$ the neutron Fermi energy reaches a point at which it becomes energetically favourable to replace neutrons by $\Lambda$ hyperons, as long as only $\Lambda N$ two-body forces are employed\cite{DSW2010,LLGP2015} . Then, however, the EoS becomes far too soft and misses the $2\,M_\odot$ constraint for the neutron star mass. 

Ongoing investigations suggest a possible way of resolving this puzzle. The starting point is an extension to chiral $SU(3)\times SU(3)$ meson-baryon EFT and the construction of interactions now involving the complete baryon and pseudoscalar meson octets. $SU(3)$ breaking effects are incorporated through the physical mass differences within the multiplets. Hyperon-nucleon interactions have been constructed in this scheme at next-to-leading order (NLO) \cite{Haidenbauer2013}. Unfortunately, the statistical quality of the existing empirical hyperon-nucleon scattering data is still too limited to warrant more detailed studies beyond NLO. There is an obvious need for a much improved hyperon-nucleon data base. 

\begin{figure}
	 \centerline{\includegraphics[width=9cm] {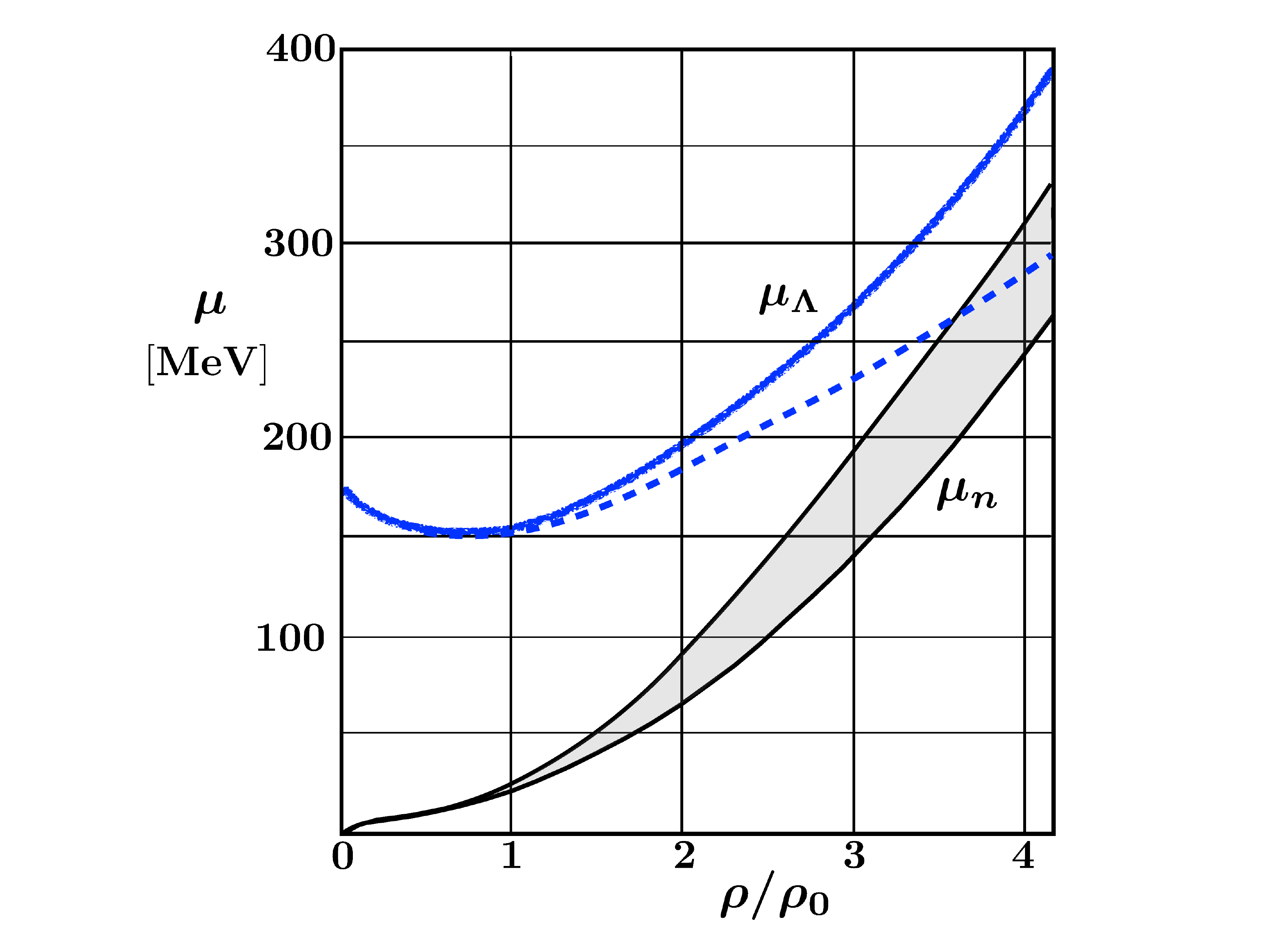}}
	\caption{Chemical potentials of $\Lambda$ hyperon and neutron, $\mu_\Lambda$ and $\mu_n$, in neutron matter as a function of baryon density (in units of normal nuclear matter density). The neutron chemical potential refers to a chiral FRG calculation with input symmetry energies in the range 30-34 MeV. The $\Lambda$ chemical potential is derived from a Brueckner calculation \cite{HMKW2017} of the $\Lambda$ single-particle potential in neutron matter. Dashed curve: $\Lambda N$ two-body interactions only. Solid curve: including $\Lambda NN$ three-body forces \cite{Petschauer2016,Petschauer2017}. }
	\label{fig:7}
\end{figure}

Such chiral $SU(3)$ calculations indicate strong $\Lambda N \rightarrow \Sigma N$ coupled-channels effects in combination with repulsive short-distance dynamics. Together these effects work to raise the onset condition for the $\Lambda$ chemical potential, $\mu_\Lambda = \mu_n$, towards higher densities.  Perhaps more significantly, $\Lambda NN$ three-body forces \cite{Petschauer2016,Petschauer2017} as they emerge from chiral $SU(3)$ EFT introduce additional repulsion\cite{HMKW2017} that raises $\mu_\Lambda$ further with increasing density. Detailed studies are now performed combining these repulsive effects in order to explore whether the condition $\mu_\Lambda = \mu_n$ can still be met in neutron stars. 

An preliminary impression of how repulsive two- and three-body $\Lambda$-nuclear interactions work against the appearance of hyperons in neutron star matter is shown in Fig.\,\ref{fig:7}. This figure is based on a Brueckner calculations of the density-dependent single-particle potential of a $\Lambda$ in neutron matter \cite{HMKW2017}. Once repulsive $\Lambda NN$ three-body forces are turned on, it is likely that the $\Lambda$ chemical potential does not meet the neutron chemical potential any more at densities relevant to the inner region of neutron stars. Such a mechanism would maintain the stiffness of the EoS and the high pressures needed to fulfill astrophysics constraints.

\section{Concluding remarks}

The focus in this presentation has been on guiding principles leading from QCD symmetries and symmetry breaking patterns to strongly interacting complex systems such as nuclei and dense baryonic matter. It indeed turns out that chiral symmetry and its spontaneous breakdown in conjunction with the confining and scale-invariance breaking QCD forces provide the basis for constructing effective field theories of low-energy QCD that lead a long way towards the understanding of nuclear forces, nuclear many-body systems and even baryonic matter under more extreme conditions such as they are encountered in the cores of neutron stars. 

In this context, a significant outcome from a non-perturbative framework using functional renormalization group methods concerns the appearance of phase transitions in the equation-of-state of cold baryonic matter. While the empirically established first-order liquid-gas transition in nuclear matter is well reproduced, strong fluctuations beyond mean-field approximation prevent a first-order chiral phase transition even at densities as high as those encountered in the core of neutron stars. In such a scheme the quest for the emergence of quark-hadron continuity and the transition to freely floating quarks in cold and compressed baryonic matter is passed over to even more extreme density scales.

\end{document}